\documentclass[twocolumn,floatfix,aps,prd,showpacs,
amsmath,amssymb]{revtex4}
\usepackage{graphicx}
\usepackage{dcolumn}
\usepackage{bm}
\usepackage{subfigure}

\def\agt{\mathrel{\raise.3ex\hbox{$>$}\mkern-14mu\lower0.6ex\hbox{$\sim$}}}
\def\alt{\mathrel{\raise.3ex\hbox{$<$}\mkern-14mu\lower0.6ex\hbox{$\sim$}}}
\def\eps{\epsilon}
\def\beq{\begin{equation}}
\def\eeq{\end{equation}}
\def\bsubeq{\begin{subequations}}
\def\esubeq{\end{subequations}}
\def\Re{{\rm Re}}
\def\Im{{\rm Im}}
\def\pd{{\partial}}
\def\rmb{{\rm b}}

\begin{document}

\title{Quasinormal ringing of acoustic black holes \\ in Laval nozzles: Numerical simulations}

\author{Satoshi Okuzumi}
\author{Masa-aki Sakagami}
\affiliation{
Graduate School of Human and Environmental Studies, Kyoto University, Kyoto,
 606-8501, Japan
}
\begin{abstract}
Quasinormal ringing of acoustic black holes in Laval nozzles is discussed.
The equation for sounds in a transonic flow is written into a Schr\"{o}dinger-type 
equation with a potential barrier, and the quasinormal frequencies are calculated semianalytically.
From the results of numerical simulations, it is shown that the quasinormal modes are actually
excited when the transonic flow is formed or slightly perturbed, as well as in the real black hole case.
In an actual experiment, however, the purely-outgoing boundary condition will not be satisfied
at late times due to the wave reflection at the end of the apparatus,
and a late-time ringing will be expressed as a superposition of ``boxed" quasinormal modes.
It is shown that the late-time ringing damps more slowly than the ordinary quasinormal
ringing, while its central frequency is not greatly different from that of the ordinary one. 
Using this fact, an efficient way for experimentally detecting the quasinormal ringing of an acoustic black hole is discussed.
\end{abstract}
\pacs{04.70.-s}

\maketitle

\section{Introduction}
When the geometry around a black hole is slightly perturbed,
a characteristic ringdown wave is emitted.
This kind of phenomenon is known as {\it quasinormal ringing},
which is expressed as a superposition of {\it quasinormal modes}~\cite{Nollert99}.
The central frequencies and the damping times of quasinormal modes
are determined by the geometry around the black hole, and thus
the gravitational quasinormal ringing of a black hole is expected 
to play the important role of 
connecting gravitational-wave observation to astronomy.

Although quasinormal modes are themselves linear perturbations, 
they are in many cases excited after {\it nonlinear} evolution of a black hole,
such as the black hole formation after the merger of binary neutron stars.
Therefore, in order to fully understand the emission process of quasinormal ringing
in such a situation, one has to resort to numerical relativity, which needs extremely powerful computational resources. 

Interestingly, an alternative way to study the physics of  
black holes has been proposed.
In a transonic fluid flow, sound waves can propagate from the subsonic region to the
supersonic region, but cannot in the opposite direction. 
Therefore, the sonic point of transonic flow can be considered as the 
``event horizon'' for sound waves, and the supersonic region as the ``black hole region''.
For this reason, a transonic fluid flow is sometimes called an {\it acoustic black hole}
\cite{Unruh81,Visser98}.
Furthermore, it is shown that the wave equation for sound waves in an inhomogeneous flow is
precisely equivalent to the wave equation for a massless scalar field in a curved spacetime~\cite{Unruh81}.
This implies that an acoustic black hole also has the quasinormal modes,
and that one can study the physics of quasinormal ringing  
without resorting to numerical relativity, ultimately {\it in  laboratories}.
Although the quasinormal modes of acoustic black holes have been studied in some recent works~\cite{Berti04,Cardoso04b,Nakano05},
it was still an open issue whether they are actually excited and, if so, how they can be excited.

In this paper, we show some results of our numerical simulations to
prove that the quasinormal ringing of sound waves is actually emitted from acoustic black holes
in some situations, just as a real black hole emits the quasinormal ringing of gravitational waves. 
For future experiments in laboratories, we treat one of the most accessible models of acoustic black holes:
a transonic flow of air in a {\it Laval nozzle}~\cite{Sakagami02,Furuhashi06}.
A Laval nozzle is a wind tunnel that is narrow in the middle, and is widely used to 
accelerate a fluid supersonically.
With this nozzle, we can easily materialize a stable transonic flow in a laboratory, just by
exerting a sufficiently large pressure difference between both ends of the nozzle.
In addition, it is known that the sonic point is always formed at the {\it throat}, where the nozzle radius
becomes the smallest. Thus, we do not need to search for the position of the sonic point
in experiments.

A quasinormal mode is characterized by a complex frequency (called {\it quasinormal frequency}) $\omega_{n}~(n=0,1,2,\cdots)$,
or equivalently, a pair of the central frequency $\Re(\omega_{n})$ 
and the quality factor $Q_n \equiv \Re(\omega_{n})/2|\Im(\omega_{n})|$.
The quality factor is a quantity that is propotional to the number of cycles of the oscillation within the damping time.
In experiments, a damping oscillation like a quasinormal ringing is inevitably buried in noise within a few damping times.
Therefore, for detecting the quasinormal ringing efficiently, it is important to design a Laval nozzle which yields a quasinormal mode with a{\it large quality factor}. 

This paper is organized as follows:
In Sec.~II, we derive the Schr\"{o}dinger-type wave equation for sounds, which
describes the ``curvature scattering" of sound waves in an acoustic black hole,
and compute the quasinormal frequencies of transonic airflow in Laval nozzles with known semianalytic methods.
In Sec.~III, we show the results of our fully-nonlinear numerical simulations to prove that 
the quasinormal modes of are actually excited in two different types of situations.
In Sec.~IV, we discuss how the quasinormal ringing changes when any outgoing waves are reflected at the upstream end of the apparatus. 
Sec.~V is devoted to our conclusion.

\section{Quasinormal Modes of Acoustic Black Holes in Laval Nozzles}
\subsection{The Schr\"{o}dinger-type wave equation}
In what follows, we assume that the radius of the Laval nozzle $r(x)$ varies sufficiently slowly along the axis of the nozzle ($x$-direction) and the airflow in the nozzle can be considered as quasi-one-dimensional. In addition, we neglect viscosity and heat conduction throughout. 

The basic equations for quasi-one-dimensional flow of a perfect fluid are
\begin{gather}
\pd_t(\rho A) + \pd_x(\rho vA) = 0 \,,
\label{eq:cont} \\
\pd_t(\rho vA) + \pd_x[(\rho v^2 + p)A] = 0 \,,
\label{eq:momentum} \\
\pd_t(\eps A) +  \pd_x[(\eps+p)vA] = 0 \,,
\label{eq:energy}
\end{gather}
where $\rho$ is the density, $v$ is the velocity, $p$ is the pressure, 
$A$ is the cross section of the nozzle, and
\beq
\eps = \frac{1}{2}\rho v^2 + \frac{p}{\gamma-1}
\eeq 
is the energy density and $\gamma=1.4$ is the heat capacity ratio for air.
Eqs.~\eqref{eq:cont},~\eqref{eq:momentum}, and~\eqref{eq:energy} are the equation of continuity,
the momentum equation, and the energy equation, respectively.
Insted of Eq.~\eqref{eq:momentum}, we can use Euler's equation
\begin{gather}
\rho(\pd_t + v \pd_x)v = -\pd_x p \,,
\label{eq:Euler}
\end{gather}
which is derived from Eqs.~\eqref{eq:cont} and~\eqref{eq:momentum}.

If the flow in the region under consideration has no entropy discontinuities (shock waves,
contact surfaces, etc.), Eq.~\eqref{eq:energy} can be replaced by the isentropic relation
\beq
p\propto\rho^\gamma \,.
\label{eq:isentropic}
\eeq
Using this relation, Eq.~\eqref{eq:Euler} is reduced to Bernoulli's equation
\beq
\pd_t\Phi + \frac{1}{2}(\pd_x\Phi)^2 + h(\rho) = 0 \,,
\label{eq:Bernoulli}
\eeq
where $h(\rho) = \int\rho^{-1}dp$ is the specific enthalpy and $\Phi = \int v\,dx$ is 
the velocity potential.

Now we derive the linearized wave equation for sounds.
To begin with, we replace $(\rho,\Phi)$ with the sum of 
the background part $(\bar\rho,\bar\Phi)$ and the perturbation part $(\delta\rho,\phi)$:
\begin{align}
\rho &= \bar\rho + \delta\rho \,, \qquad \bar\rho \gg |\delta\rho|   \,, 
\label{eq:split_rho} \\
\Phi &= \bar\Phi + \phi \,,       \qquad |\pd_x\bar\Phi| \gg |\pd_x\phi| \,,
\label{eq:split_Phi}
\end{align}

Substituting Eqs.~\eqref{eq:split_rho} and \eqref{eq:split_Phi} into 
Eqs.~\eqref{eq:cont} and \eqref{eq:Bernoulli}, and eliminating $\delta\rho$, 
we obtain the wave equation for the velocity potential perturbation $\phi$,
\beq
\Bigl[\Bigl( \pd_t + \pd_x\bar{v} \Bigr)\frac{\bar{\sigma}}{c_s^2}
\Bigl( \pd_t + \bar{v}\pd_x \Bigr)-
\pd_x(\bar{\sigma}\partial_x)\Bigr] \phi = 0\;,
\label{eq:wave1}
\eeq
where $\bar{\sigma} = \bar{\rho}A$, $\bar{v}=\pd_x\bar{\Phi}$, and $c_s = \sqrt{dp/d\rho} = \sqrt{\gamma p(\bar{\rho})/\rho}$. Note that the sound speed $c_s$ depends on the local background state, and is therefore a function of $x$. 
For $A\equiv{\rm const.}$, Eq.~\eqref{eq:wave1} is reduced to the one-dimensional version of Unruh's 
wave equation~\cite{Unruh81}.
In the following, we omit the bars over the background quantities. 

Next, we consider the stationary background and Fourier transform $\phi(t,x)$
in terms of $t$:
\beq
\phi_\omega(x) = \int\phi(t,x)e^{i\omega t}dt \,.
\eeq
Then, Eq.~\eqref{eq:wave1} gives the time-independent differntial equation for $\phi_\omega$,
\begin{gather}
\Biggl[
\frac{d^2}{dx^2}+\frac{1}{\sigma(1-M^2)}\Bigl[\frac{d}{dx}		\Bigl(\sigma(1-M^2)\Bigr)+2i\omega\frac{\sigma v}{c_s^2}\Bigr]\frac{d}{dx} \nonumber \\[-3pt]
 + \frac{1}{\sigma(1-M^2)}\Bigl[i\omega\frac{d}{dx}\Bigl(\frac{\sigma v}{c_s^2}\Bigr) + \omega^2\frac{\sigma}{c_s^2} \Bigr]
\Biggr]\phi_\omega = 0, \label{eq:wave1D-2}
\end{gather}
where $M=|v|/c_s$ is the Mach number.
In order to transform Eq.~\eqref{eq:wave1D-2} into the Schr\"{o}dinger-type equation, we introduce a new function $H_\omega(x)$ and the ``tortoise coordinate"  $x^*$ by 
\begin{gather}
H_\omega(x) = g^{1/2}\int dt~e^{i\omega[t-f(x)]}\phi(t,x), \\
x^* = c_{s0}\int\frac{dx}{c_s(1-M^2)},
\end{gather}
where
\begin{gather}
g = \frac{\sigma}{c_s}, \\
f(x) = \int\frac{|v|\,dx}{c_s^2-v^2},
\end{gather}
and $c_{s0}$ is the stagnation sound speed, constant over the isentropic
region.
Then, Eq.~\eqref{eq:wave1D-2} can be rewritten as
\beq
\biggl[ \frac{d^2}{dx^{*2}} + \kappa^2 - V(x^*) \biggr] H_\omega = 0, \label{eq:Sch1}
\eeq
where
\begin{gather}
\kappa = \frac{\omega}{c_{s0}}, \\
V(x^*) = \frac{1}{g^2}\biggl[\; \frac{g}{2}\frac{d^2g}{dx^{*2}} 
    - \frac{1}{4}\Bigl(\frac{dg}{dx^*}\Bigr)^2 \;\biggr].
\end{gather}
The effective potential $V(x^*)$ has the dimension of (length)$^{-2}$ and 
characterizes the ``curvature scattering'' of sound waves on the acoustic black hole. 
In a Laval nozzle, all of the profiles of the nondimensional quantities ($\rho/\rho_0$, $p/p_0$, $M$, etc.) for the transonic flow are uniquely determined 
by a function $A(x^*)/A_{\rm th}$,
where $A_{\rm th}$ is the cross-sectional area at the throat of the nozzle~\cite{LiepmannRoshko}.
Hence, the structure of the effective potential $V(x^*)$ is also completely determined by
$A(x^*)/A_{\rm th}$ only. 

\begin{figure*}
\centering
\includegraphics[width=17cm]{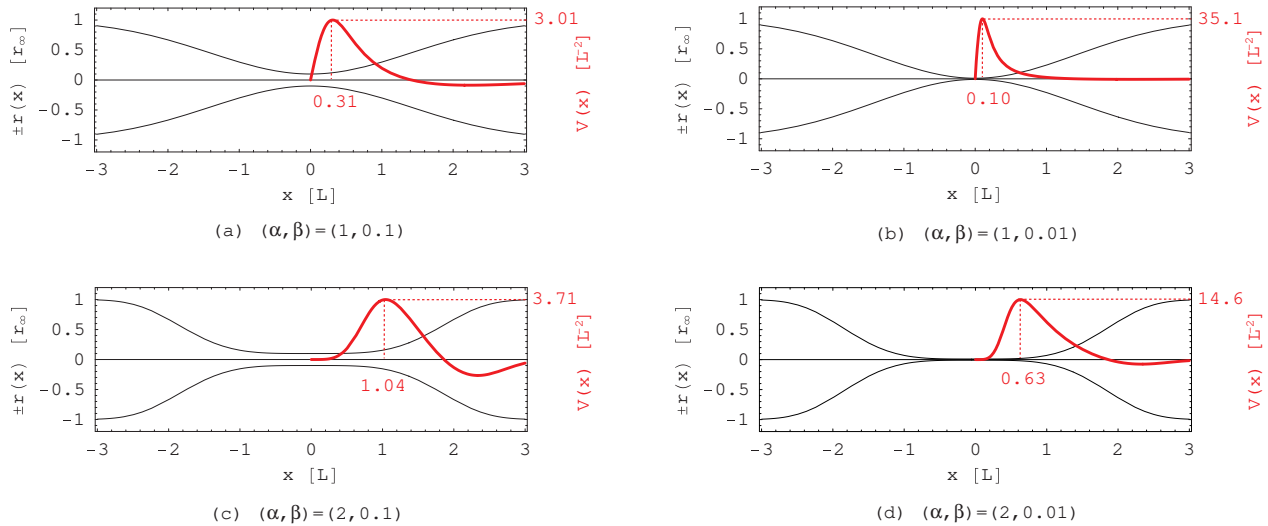}
\caption{
Forms of different Laval nozzles $\pm r(x)$ (thin curves) and the potentials $V(x^*)$ (thick curves) 
for stationary transonic flow, as functions of $x$.
The fluid is assumed to flow in the negative $x$-direction, and thus $x<0$ is the 
supersonic region.
Also indicated are the maximum point and the height of each $V(x^*)$ by the vertical and the 
horizontal dotted line, respectively.
}
\label{fig:nozzle}
\end{figure*}

In this study, we consider a family of Laval nozzles which have the following form:
\begin{align}
A(x) &= {\pi} \, r(x)^2 \; , 
\label{eq:A(x)}\\
r(x) &= r_\infty - r_\infty (1-\beta) \, \mathrm{exp}[-(x/2L)^{2\alpha}]\; ,
\label{eq:r(x)}
\end{align}
where $\alpha$ is a positive integer and $\beta \equiv r(0)/r_\infty \in (0,1)$.
This type of Laval nozzle has a throat at $x=0$ and two, infinitely long ``tanks'' 
with a constant radius $r_\infty$ at $x \gg L$ and $x \ll -L$ . 
Note that the values of $r_\infty$ and $L$ do not affect the structure of the 
nondimensionalized potential $V(x^*)L^2$, and may be taken 
so that the quasi-one-dimension approximation ($dr/dx \ll 1$) holds.
Fig.~\ref{fig:nozzle} shows four examples of the form of the nozzle $\pm r(x)$ and the potential barrier $V(x^*)$ for the stationary transonic flow. 

\subsection{Evaluation of quasinormal frequencies}
Having obtained the form of the potential $V(x^*)$, we can compute the quasinormal
frequencies by solving the Schr\"{o}dinger-type wave equation (\ref{eq:Sch1}) 
with the outgoing boundary conditions,
\begin{align}
H_\omega (x^*) &\propto e^{-i\kappa x^*} \,, \qquad x^* \to -\infty \,,
\label{eq:BC-} \\ 
H_\omega (x^*) &\propto e^{+i\kappa x^*} \,, \qquad x^* \to +\infty \,.
\label{eq:BC+}
\end{align} 

In the case of a real black hole, the functional form of the effective potential
is known analytically, and thus applicable is Leaver's method~\cite{Leaver85}, which
expands the wave function into a certain series and solves the continued fraction equation for the expansion coefficents.
In the case of our acoustic black hole, however, the form of the potential 
can be known only numerically, and thus we have to apply an another method.

In this study, we adopt the P\"{o}schl-Teller potential approximation~\cite{Mashhoon84} and the WKB approximation.

\subsubsection{P\"{o}schl-Teller potential approximation}
This method approximates a potential barrier with the P\"{o}schl-Teller potential, 
for which the scattering problem can be exactly solved. 
The P\"{o}schl-Teller potential $V_{\rm PT}(x^*)$ is defined by
\beq
V_{\rm PT}(x^*) = \frac{V_0}{\cosh^2[K(x^*-x^*_0)]} \\,
\eeq
where $x_0^*$ is the position of the potential peak, and $V_0 = V_{\rm PT}(x^*_0)>0$ and $-2K^2 = V_{\rm PT}''(x^*_0)/V_0<0$ are the height and the curvature of the potential, respectively. 
Eq.~\eqref{eq:Sch1} with $V(x^*)=V_{\rm PT}(x^*)$ has a solution $H_\omega(x^*) \propto h_\omega(x^*)$ which behaves at $x^*\to\pm\infty$
as
\begin{align}
h_\omega
&\sim
e^{-i\kappa(x^*-x^*_0)} \,, & x^*\to-\infty \,, 
\label{eq:asympt-} \\
h_\omega
&\sim
A(\kappa)e^{-i\kappa(x^*-x^*_0)} + B(\kappa)e^{i\kappa(x^*-x^*_0)}\,,& x^*\to+\infty \,, 
\label{eq:asympt+}
\end{align}            
where $A(\kappa)$ and $B(\kappa)$ are defined by
\begin{align}
A(\kappa) &= \frac{\Gamma(1-i\kappa/K)\Gamma(-i\kappa/K)}
                 {\Gamma(s_+ -i\kappa/K)\Gamma(s_- -i\kappa/K)} \,,  \\
B(\kappa) &= \frac{\Gamma(1-i\kappa/K)\Gamma(i\kappa/K)}
                 {\Gamma(s_+)\Gamma(s_-)} \,, \\
s_\pm &= \frac{1}{2}\biggl[1\pm i \left(\frac{4V_0}{K^2}-1\right)^{1/2}\biggr] \,.         
\end{align}
Comparing Eqs.~\eqref{eq:BC-} and \eqref{eq:BC+} with Eqs.~\eqref{eq:asympt-} and \eqref{eq:asympt+},
$h_\omega(x^*)$ is found to be the quasinormal mode of the P\"{o}schl-Teller potential 
if $A(\kappa)=0$, or
\begin{align}
\kappa = \kappa_n &\equiv -iK(s_+ + n) \notag \\
&= \Bigl(V_0 - \frac{K^2}{4} \Bigr)^{1/2} - iK\Bigl(n+\frac{1}{2}\Bigr) \,,
\label{eq:QNM_PT}
\end{align}
where $n$ is a non-negative integer.
With this relation, we can evaluate the quasinormal frequencies of a {\it general} potential $V(x^*)$ 
by approximating the form of the potential around $x^*=x^*_0$ with the P\"{o}schl-Teller potential,
i.e., by setting $V_0 = V(x^*_0)$ and $-2K^2 = V''(x^*_0)/V(x^*_0)$ in Eq.~\eqref{eq:QNM_PT}.
In general, this method gives an acurrate quasinormal frequency for small $n$,
since such a mode is basically determined by the structure of $V(x^*)$ at the
maximum. 

\subsubsection{WKB approximation}
This method constructs the quasinormal mode solutions by approximating $H_\omega(x^*)$ with WKB functions in both sides of the potential barrier, matching them across the potential peak, and imposing the outgoing boundary conditions on it.
The first-order WKB formula for quasinormal frequencies was derived by Schutz and Will \cite{SchutzWill85},
and the higher-order versions have been developed by Iyer and Will \cite{IyerWill87}, and Konoplya \cite{Konoplya0304}.

The WKB formula for $\kappa_n$ reads
\begin{align}
\kappa_n^2 = &V(x^*_0) -i\Bigl(n+\frac{1}{2}\Bigr)[-2V''(x^*_0)]^{\frac{1}{2}} \notag \\
           &+ (\textrm{higher-order correction terms})\;,
\label{eq:QNM_WKB}
\end{align}
where the higher-order correction terms depend on $V^{(\geq3)}(x^*_0)$ and $n$ (see Ref.~\cite{IyerWill87,Konoplya0304}).
This method also gives an acurrate quasinormal frequency for small $n$.

\subsubsection{Results}

\begin{table}
\centering
\begin{ruledtabular}
\begin{tabular}{llcccc}
      &         & \multicolumn{4}{c}{$\omega_0\;[c_{s0}/L]$} \\ \cline{3-6}
{\raisebox{1.5ex}[0pt]{$\alpha$}}&\multicolumn{1}{c}{{\raisebox{1.5ex}[0pt]{$\beta$}}} &(PT)&(WKB1)&(WKB2)&(WKB3) \\ \hline
    1 & $0.1$   & $1.29 - 1.16i$ &                &                &               \\%
	  & $0.03$  & $2.68 - 2.04i$ &                &                &               \\%
	  & $0.01$  & $4.81 - 3.46i$ &                &                &               \\%
	  & $0.003$ & $8.73 - 6.47i$ &                &                &               \\ \hline
    2 & $0.1$   & $1.68 - 0.95i$ & $2.10 - 0.85i$ & $1.78 - 1.01i$ & $1.73 - 0.91i$\\%
      & $0.03$  & $2.63 - 1.06i$ & $3.00 - 0.99i$ & $2.65 - 1.11i$ & $2.61 - 1.02i$\\
      & $0.01$  & $3.60 - 1.28i$ & $4.00 - 1.20i$ & $3.57 - 1.34i$ & $3.54 - 1.26i$\\%
      & $0.003$ & $4.94 - 1.66i$ & $5.44 - 1.57i$ & $4.86 - 1.76i$ & $4.83 - 1.68i$\\ \hline
    3 & $0.1$   & $2.23 - 1.12i$ & $2.70 - 1.02i$ & $2.43 - 1.14i$ & $2.46 - 1.19i$\\
      & $0.03$  & $3.22 - 1.05i$ & $3.53 - 1.00i$ & $3.35 - 1.06i$ & $3.38 - 1.14i$\\
      & $0.01$  & $4.07 - 1.09i$ & $4.34 - 1.11i$ & $4.13 - 1.11i$ & $4.15 - 1.16i$\\
      & $0.003$ & $5.07 - 1.24i$ & $5.35 - 1.20i$ & $5.08 - 1.26i$ & $5.09 - 1.29i$\\ \hline
    4 & $0.1$   & $2.76 - 1.45i$ & $3.36 - 1.25i$ & $3.09 - 1.36i$ & $3.15 - 1.48i$\\
      & $0.03$  & $3.92 - 1.18i$ & $4.24 - 1.14i$ & $4.10 - 1.17i$ & $4.12 - 1.23i$\\%
      & $0.01$  & $4.77 - 1.12i$ & $5.02 - 1.10i$ & $4.86 - 1.13i$ & $4.85 - 1.09i$\\%
      & $0.003$ & $5.69 - 1.09i$ & $5.91 - 1.15i$ & $5.69 - 1.20i$ & $5.66 - 1.09i$\\ 
\end{tabular}
\end{ruledtabular}
\caption{
Frequencies  $\omega_0$ and quality factors $Q_0=\Re(\omega_0)/2\Im(\omega_0)$ of the least-damped quasinormal modes for different Laval nozzles,
evaluated with the P\"{o}schl-Teller approximation method (PT) and the WKB method up to the third-order
(WKB1-3).
The WKB results for $\alpha=1$ are not listed, since they do not converge as the WKB order increases. 
}
\label{table:QNMs}
\end{table}
Table~\ref{table:QNMs} shows the least-damped quasinormal frequencies $\omega_0=c_{s0}\kappa_0$ 
for Laval nozzles with different $(\alpha,\beta)$, obtained by the P\"{o}schl-Teller potential approximation and the WKB approximation up to the third WKB order.
For $\alpha \geq 2$, the higher-order correction terms in Eq.~\eqref{eq:QNM_WKB} converges well, 
and the P\"{o}schl-Teller values agree with
the third order WKB values within $\sim10\%$ accuracy.
For $\alpha=1$, on the other hand, the correction terms are found to diverge as the WKB order increases,
meaning that the WKB approximation breaks down.
There is a possibility that the classical turning points across which the WKB functions are mathced are too distant from each other for these nozzles. 
To avoid confusion, we do not list the WKB values for $\alpha=1$ in the Table. 

We find that $\Re(\omega_0),\Im(\omega_0) \sim c_{s0}/L$, meaning that the wavelength and the damping
length of the least damped mode are both determined by the curvature radius ($\sim L$) of the Laval nozzle.
We also find that the quality factor $Q_0=\Re(\omega_0)/|2\Im(\omega_0)|$ of the least-damped quasinormal mode is $0.5\alt Q_0 \alt 3 \ll 10$,
which implies that the quasinormal ringing damps out for only a few oscillation cycles.

\subsection{Numerical Simulations}

\subsubsection{Settings}
In the numerical simulations, a Laval nozzle is assumed to have a form described by 
Eqs.~\eqref{eq:A(x)} and \eqref{eq:r(x)},
and the flow in the nozzle is treated as quasi-one-dimensional.
In order to solve Eqs.~\eqref{eq:cont},~\eqref{eq:momentum}, and~\eqref{eq:energy} numerically, 
we adopted the MUSCL-Hancock scheme~\cite{Toro},
the second-order version of the Godunov scheme~\cite{Godunov59,vanLeer77}.
This scheme enables us to resolve shock waves generated in the flow in a good accuracy.
The computational domain is set to $-7L \leq x \leq 7L$, divided into 800-1000 computing cells.
 
In the case of an astrophysical black hole, quasinormal ringing appears when it is
formed or when a test particle falls into it.
As their analogues, we consider two types of situation for our acoustic black holes:
(I)\textit{acoustic black hole formation} and (II)\textit{weak-shock infall}. 
In simulations of type I, the initial state is set to be stagnant and homogeneous throughout the fluid,
$(\rho(x),v(x),p(x))_{t=0}\equiv(\rho_0,0,\,p_0)$.
At $t=0$, the fluid in the nozzle begins to be ``pumped out" from one end, 
at a rate high enough for the transonic flow to be formed.
Fluid is allowed to enter freely from the other end so that the transonic flow can be stationary.
In simulations of type II, on the other hand, flow is stationary and transonic from the first, 
and a weak shock wave is injected from the upper end to disturb the flow in the nozzle slightly.  More infomation on the setting is summarized in Table~\ref{table:settings}.
\begin{table*}
\centering
\begin{ruledtabular}
\begin{tabular}{ccccc}
  &  & \multicolumn{2}{c}{boundary condition} &  \\ \cline{3-4}
   {\raisebox{1.5ex}[0pt]{type}}   &  {\raisebox{1.5ex}[0pt]{initial state}}   &  downstream ($x = -7L$)      &  upstream ($x = 7L$) & {\raisebox{1.5ex}[0pt]{final state}}  \\ \hline      
 I & $(\rho,v,p)=(\rho_0,\,0,\,p_0)$ &  $(\rho,v,p)=(1.0\rho_0,\,-0.1c_{s0},\,0.7p_0)$ &  transmissive & stationary transonic\\
 II  &  stationary transonic & transmissive & transmissive & stationary transonic \\
\end{tabular}
\end{ruledtabular}
\caption{Settings of type I and type II simulations. $\rho_0$, $c_{s0}$, and $p_0$ 
denote the stagnation density, sound speed, and pressure of the flow, respectively.
}
\label{table:settings}
\end{table*}
\subsubsection{Results}
\begin{figure}
\includegraphics[width=8.5cm]{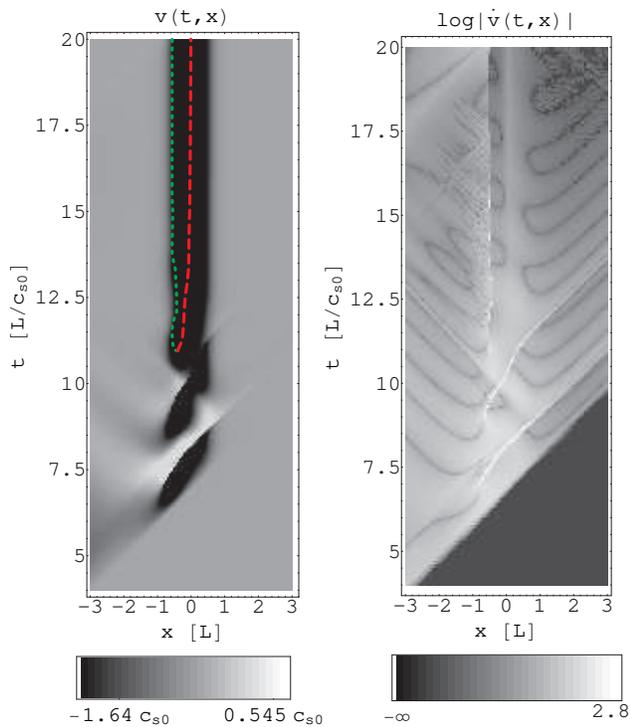}
\caption{
Spacetime distribution of $v$ (left) and $\log|\dot{v}|$ (right)
obtained from a numerical simulation of type I.
The nozzle parameters are set to $(\alpha,\beta)=(2,0.01)$. 
The dashed curve and the dotted curve represent the position of the sonic point and the steady shock, respectively.
}
\label{fig:Evolution1}
\end{figure}
Fig.~\ref{fig:Evolution1} shows the spacetime distribution of $v$ and $\log|\dot{v}|$
obtained from the type I simulation for $(\alpha,\beta)=(2,0.01)$.
At each position $x$, the fluid begins to move towards the lower boundary at $t \simeq (x+7L)/c_{s0}$.
In the ``tank" regions ($|x| \agt L$) the flow quickly relaxes to the stationary state, while
in the ``throat" region ($|x| \alt L$) the flow settles down gradually, ejecting shock waves towards both
tank regions. This difference reflects the difference in the cross section
between the throat and the tanks; i.e., the cross section of the throat is much smaller than
that of the tanks, and thus the backreaction to the expansion of the fluid is much greater.
The sonic point is formed at $t \simeq 13L/c_{s0}$, and then approaches to the throat ($x=0$) gradually.
The discontinuity appearing  in the supersonic region
is the wavefront of a steady shock, which is peculiar to the stationary transonic flow in a Laval nozzle~\cite{LiepmannRoshko}.
In the right panel of Fig.~\ref{fig:Evolution1}, we can see a ringdown wave
emittied for $t\agt 8L/c_{s0}$ from $x \simeq L$, around which the 
effective potential barrier is to be formed [see Fig.~\ref{fig:nozzle}(d)].
This implies that the quasinormal modes are actually excited 
during the transonic flow formation.

\begin{figure}
\centering
\includegraphics[width=8.0cm]{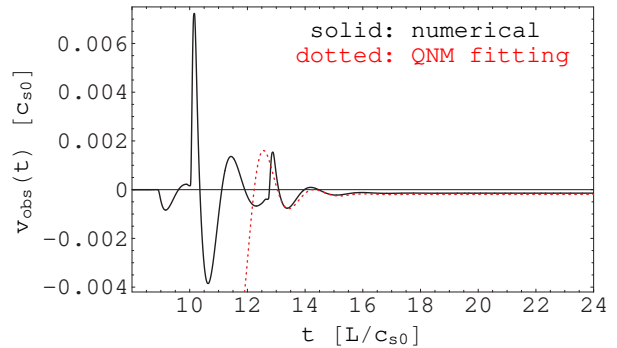}
\caption{
Numerical waveform $v_{\rm obs}(t) = v(t,2.0L)$ (solid curve) obtained from the same simulation as that shown in Fig.~\ref{fig:Evolution1}, compared to the least-damped quasinormal mode (dotted curve) with the frequency $\omega_0 = (3.60 - 1.28i)\,c_{s0}/L$.
}
\label{fig:fitting1}
\end{figure}
In Fig.~\ref{fig:fitting1}, the numerical waveform of $v$ observed at $x=2.0L$ is plotted.
The bumps at $t \simeq 10L/c_{s0}$ and $13L/c_{s0}$ are due to the passing of shock waves generated 
in the throat region.
For $t \agt 13L/c_{s0}$, we find that the waveform agrees very well with the least-damped ($n=0$)
quasinormal mode $\omega_0 = 3.60 - 1.28i$ calculated by the P\"{o}schl-Teller potential approximation.
From this fact, the ringdown wave appearing in Fig.~\ref{fig:Evolution1} is proved to be the quasinormal
ringing.
\begin{figure}
\centering
\vspace{-0.5cm}
\includegraphics[width=8.5cm]{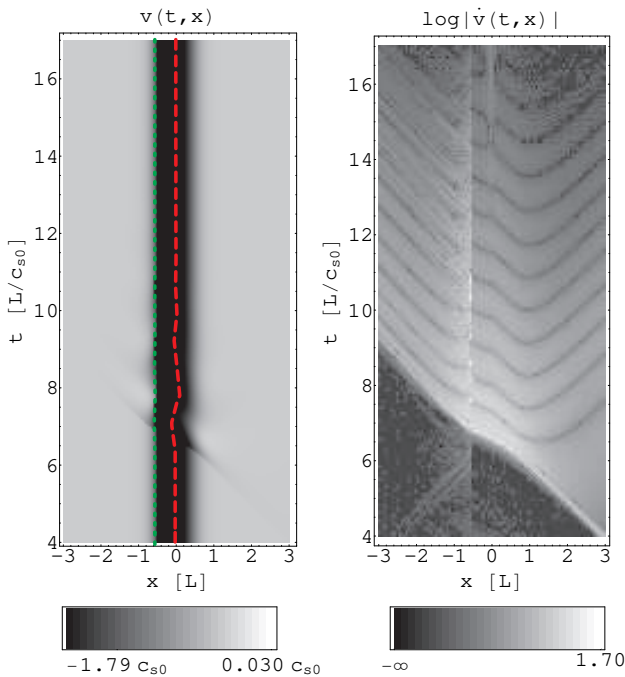}
\vspace{-3.5cm}
\caption{
Spacetime distribution of $v$ (left) and $\log|\dot{v}|$ (right)
obtained from a numerical simulation of type II, 
with nozzle parameters $(\alpha,\beta)=(2,0.01)$. 
}
\label{fig:Evolution2}
\end{figure}
\begin{figure}
\centering
\includegraphics[width=8.0cm]{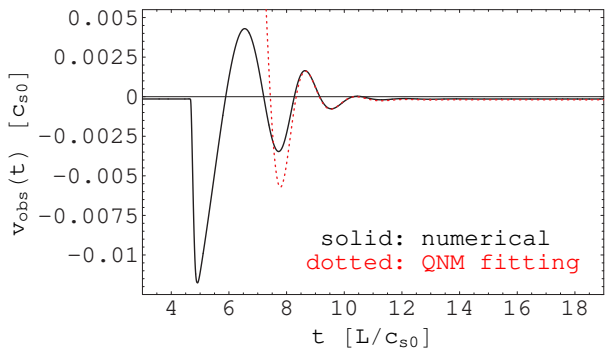}
\caption{
Numerical waveform $v_{\rm obs}(t) = v(t,2.0L)$ (solid curve) obtained from the same simulation as that shown in Fig.~\ref{fig:Evolution2}, compared to the least-damped quasinormal mode (dotted curve).
}
\label{fig:Fitting2}
\end{figure}

The quasinormal ringing of transonic flow can be observed more distinctly in type II simulations, one of whose results is shown in Figs.~\ref{fig:Evolution2} and~\ref{fig:Fitting2}. We find that the observed
waveform is again in good agreement with the least-damped quasinormal mode.

\section{Boxed Quasinormal Modes: Effect of the Wave Reflection at a Boundary}
Quasinormal modes are essentially characterized by the outgoing boundary conditions,
Eqs.~\eqref{eq:BC-} and \eqref{eq:BC+}.
In our acoustic black holes, Eq.~\eqref{eq:BC-} is always satisfied due to the presence of the acoustic horizon.
On the other hand, whether Eq.~\eqref{eq:BC+} is satisfied or not depends on 
the structure of the Laval nozzle in the upstream region.
In the simulations shown in the last Section, for example, the upstream tank is assumed to be infinitely long,
and therefore no wave is allowed to return to the potential barrier.
In a such case, Eq.~\eqref{eq:BC+} is {\it always} satisfied and the potential barrier will {\it always} emit quasinormal ringing.
In a real experimental configuration, however, the upstream tank has a {\it finite} length, 
and any outgoing wave emitted from the potential barrier is (perfectly or partially) {\it reflected} at the boundary, i.e., at the end of the tank.
In such a case, Eq.~\eqref{eq:BC+} is satisfied {\it only at early times}, that is, before the wavefront of the reflected wave arrives at the potential barrier. Hence, in real experiments, the sound wave observed {\it at late times} will be different from an ordinary quasinormal ringing. 

Now we consider how the wave reflection at the boundary alters the quasinormal ringing in the Laval nozzle. In the following, we assume that the upstream tank 
has a reflective boundary wall perpendicular to the axis of the nozzle at $x=x_c \gg x_0$.
Then, the appropriate boundary conditions for $H_\omega$ are
\begin{align}
  H_\omega &\propto e^{-i \kappa x^*}, & x^* \rightarrow -\infty, \label{eq:BC_cs-} \\ 
  H_\omega &\propto e^{+i \kappa(x^*-x^*_c)} + {\cal R}_\omega\,e^{-i\kappa(x^*-x^*_c)},
                                          & x^* \rightarrow x_c^*, \label{eq:BC_cs}
\end{align}
where ${\cal R}_\omega \in [0,1]$ is the reflection coefficient of the boundary wall.
The boundary wall is perfectly reflective ($v|_{x^*_c}\propto\pd_xH_\omega|_{x^*_c} = 0$) for ${\cal R}_\omega=1$, and partially absorbing for $0<{\cal R}<1$.
The solutions of the wave equation~\eqref{eq:Sch1} with these boundary conditions are no more quasinormal modes in a narrow sense,
but are yet expected to be ``quasinormal", with complex characteristic frequencies.
For this reason, we shall refer them to {\it ``boxed" quasinormal modes}
after Refs.~\cite{Cardoso04a,Cardoso04b}.

\subsection{Evaluation of boxed quasinormal frequencies}
In order to calculate the frequencies of boxed quasinormal modes, we approximate $V(x^*)$ with the P\"{o}schl-Teller potential used in Sec~II.
Comparing the upstream boundary condition~\eqref{eq:BC_cs} with the asymptotic form~\eqref{eq:asympt+} of the particular solution $h_\omega(x^*)$,
we obtain the characteristic equation for $\kappa$,
\beq
 \frac{\Gamma(s_+)\Gamma(s_-)\Gamma(-i\kappa/K)}{\Gamma(s_+-i\kappa/K)\Gamma(s_--i\kappa/K)\Gamma(i\kappa/K)}
= {\cal R}_\omega e^{2i\kappa\Delta} \,,
\label{eq:BQNM_PT}
\eeq
where $\Delta = x^*_c - x^*_0$ is the distance between the boundary wall and the peak of the potential.
For simplicity, we assume the reflection coefficent ${\cal R}_\omega$ is independent of $\omega$
and rewrite it as ${\cal R}_\omega \equiv {\cal R}$. We find a number of solutions for fixed $\Delta$, and the number is considered to be infinite for a reason mentioned below.
We refer to the solutions as $\kappa_{\rmb,m}$, where $m=0,1,2,\cdots$ are labeled in the order of increasing $\Re(\kappa_{\rmb,m})$, not of increasing $|\Im(\kappa_{\rmb,m})|$.

\begin{figure}
\centering
\subfigure
{
\hspace{4mm}
\includegraphics[width=7.5cm]{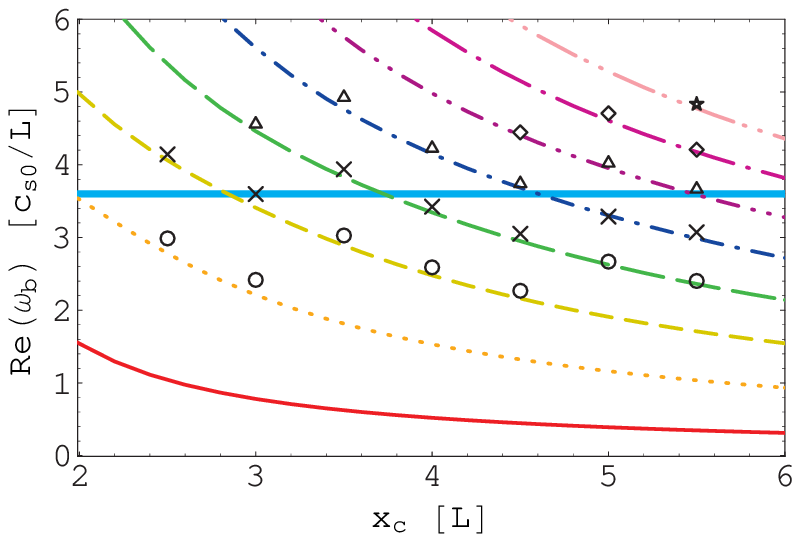}
}
\\[-0.5cm]
\subfigure
{
\includegraphics[width=7.5cm]{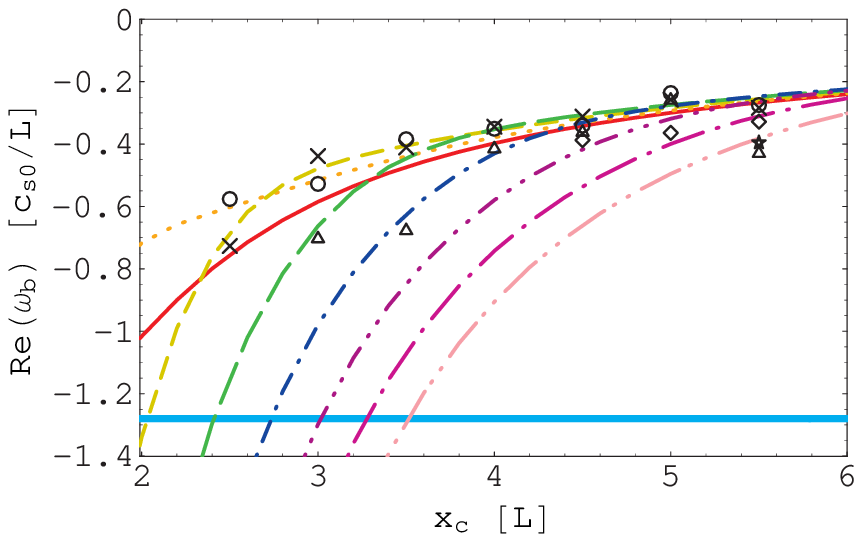}
}
\caption{
Boxed quasinormal frequencies for $(\alpha,\beta)=(2,0.01)$ and ${\cal R}=0.089$, as functions of $x_c$.
The curves indicate eight of the numerical solutions to Eq.~\eqref{eq:BQNM_PT}, $\omega_{\rmb,m} (m=0,\cdots,7)$.
For comparison, the least-damped ordinary quasinormal frequency $\omega_0$ is indicated by the thick horizontal line. 
The symbols `$\circ$', `$\times$', `$\vartriangle$', `$\diamond$', and `$\star$' represent the ``best fitting" frequencies ${\bm\omega}_{\rm min}$ for the waveforms obtained through type I' simulations (see Sec.~III.B).
}
\label{fig:CSdistanceReIm}
\end{figure}
In Fig.~\ref{fig:CSdistanceReIm}, we plot some of the numerical solutions of Eq.~\eqref{eq:BQNM_PT},
$\omega_m\;(m=0,\cdots,7)$, for $(\alpha,\beta)=(2,0.01)$ and ${\cal R}=0.089$ as functions of
$x_c\;(\simeq \Delta + 0.63L$). We find that the central frequencies $\Re(\omega_{\rmb,m})$ approximately satisfy the relation $\Re(\omega_{\rmb,m+1})-\Re(\omega_{\rmb,m}) \simeq \pi c_{s0}/\Delta$,
which is the same as the one satisfied by the normal modes in a cavity with length $\Delta$. 
This means that the boxed quasinormal modes are basically the same as the normal modes in a cavity;
the only difference is that the ``cavity'' is leaky, i.e., the boundary wall and the potential barrier
transmit sound waves partially.
For this reason, the number of the boxed quasinormal modes is considered to be infinite as is for the normal modes.

We also find that the boxed quasinormal modes with
$\Re(\omega_{\rmb,m})\alt\Re(\omega_0)$ have the damping rates $|\Im(\omega_{\rmb,m})|$ that are much smaller than that of the least-damped quasinormal modes, $|\Im(\omega_{0})|$.
From this fact, the ringing observed at late times is expected to damp much more slowly than
the quasinormal ringing at early times.
In fact, it is found from the results of the numerical simulations that the modes with $\Re(\omega_{\rmb,m})\simeq\Re(\omega_0)$ are the most strongly excited, and as a result the late-time ringing has the quality factor much smaller than that of the early-time quasinormal ringing (see Sec.~III.B).
We also calculated the frequencies for $m\geq8$ and checked that their damping rates are higher than that for $m = 7$ (at least in the range $2L<x_c<6L$).
Therefore, the modes with large $m$ will not dominate the late-time ringing, even if the number
of such modes is infinite.

\subsection{Numerical simulations}
In order to verify that the late-time waveform is acually characterized by the boxed quasinormal modes, we perform modified versions of type I simulations, to which we refer as type I' simulations. In type I' simulations, the upstream boundary is set to be reflective ($0<{\cal R}\leq1$) and located at $x=x_c(>x_0)$, while the initial state in the nozzle and the condition at the downstream boundary are the same as those of type I. A summary on the setting of type I' simulations is shown in Table~\ref{table:setting1'}.

\begin{table*}
\centering
\begin{ruledtabular}
\begin{tabular}{ccccc}
  &  & \multicolumn{2}{c}{boundary condition} &  \\ \cline{3-4}
   {\raisebox{1.5ex}[0pt]{type}}   &  {\raisebox{1.5ex}[0pt]{initial state}}   &  downstream ($x = -7L$)      &  upstream ($x = x_c$) & {\raisebox{1.5ex}[0pt]{final state}}  \\ \hline      
 I & $(\rho,v,p)=(\rho_0,\,0,\,p_0)$ &  $(\rho,v,p)=(1.0\rho_0,\,-0.1c_{s0},\,0.7p_0)$ &  reflective ($0<{\cal R}\leq 1$) & stationary transonic\\
\end{tabular}
\end{ruledtabular}
\caption{Setting of type I' simulations. ${\cal R}$ denotes the reflection coefficient at the upper boundary.
}
\label{table:setting1'}
\end{table*}
\begin{figure}
\centering
\subfigure
{
\includegraphics[width=8.0cm]{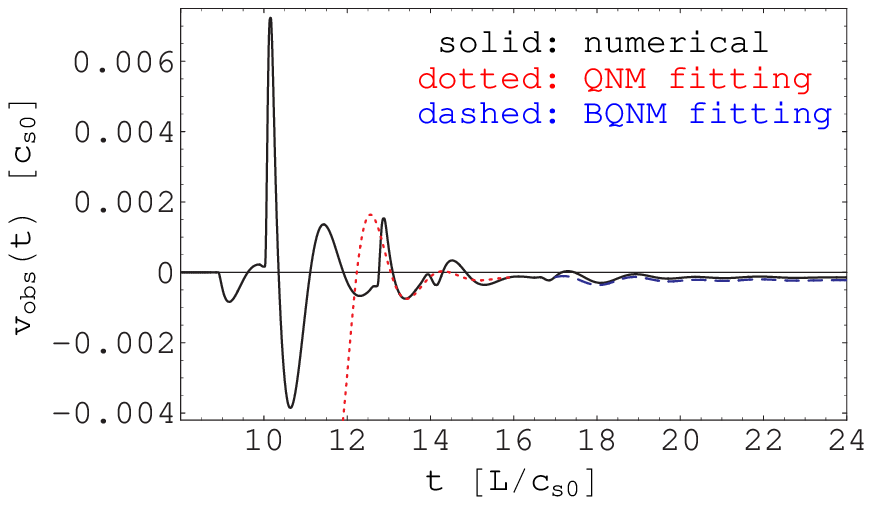}
}
\\[-0.5cm]
\subfigure
{
\includegraphics[width=8.0cm]{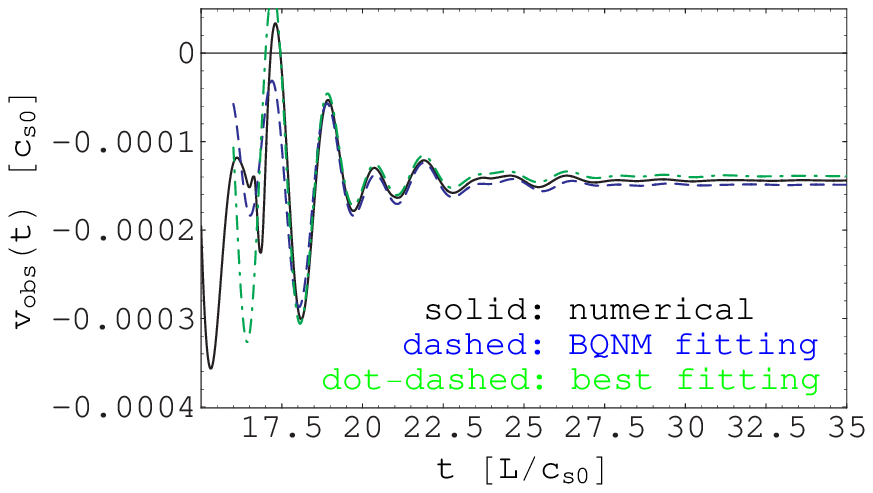}
}
\caption{
Numerical waveform $v_{\rm obs}(t) = v(t,2.0L)$ (solid) for an early time (upper panel) and a late time (lower panel), obtained from the type I' simulation with $(\alpha,\beta,x_c,{\cal R})=(2,0.01,4.0L,0.089)$.
For comparison, the least-damped quasionormal mode (dotted), a superposition
of three boxed quasinormal modes (dashed), and the ``best fitting'' function (dot-dashed) are also plotted.
}
\label{fig:fittingCS1}
\end{figure}
Fig.~\ref{fig:fittingCS1} shows the numerical waveform of $v$ at $x_{\rm obs}=2.0L$ obtained from the type I' simulation with $(\alpha,\beta,x_c,{\cal R})=(2,0.01,4.0L,0.089)$.
At early times ($t \alt 14L/c_{s0}$), the waveform looks identical to that shown in Fig.~\ref{fig:fitting1}, and is well fitted to the least-damped quasinormal mode again. 
In fact, $t=14L/c_{s0}$ is just the time at which the bump observed at $t\simeq10L/c_{s0}$ returns to
$x=x_{\rm obs}=2L$ after reflected at the boundary ($x=x_c=4L$).
At late times ($t \agt 14L/c_{s0}$), the waveform is no longer fitted to the ordinary quasinormal modes.
In order to verify that the late-time waveform is a superposition of the boxed quasinormal modes, we decompose the waveform into damped sinusoids as follows: first, we calculate the function 
\begin{gather}
E({\bm\omega},{\bm C},v_\infty) \equiv \int_{t_1}^{t_2}\Bigl[v_{\rm obs}(t)- u({\bm\omega},{\bm C},v_\infty ;t) \Bigr]^2 dt, \\ 
u({\bm\omega},{\bm C},v_\infty ;t) \equiv v_\infty + \Re\biggl[\sum_{j=1}^{j_{\rm max}} C^{(j)}e^{-i\omega^{(j)}(t-t_1)}\biggr],
\end{gather}
where ${\bm\omega}=(\omega^{(1)},\cdots,\omega^{(j_{\rm max})})$, ${\bm C}=(C^{(1)},\cdots,C^{(j_{\rm max})})$, and $v_\infty$ are fitting parameters, and $t_1,t_2$ are taken to be large enough. 
Then, we search for the set of parameters $({\bm\omega},{\bm C},v_\infty)_{\rm min}$ that minimizes $E$. 
In this example, setting $j_{\rm max}=3$, $t_1=19L/c_{s0}$, and $t_2=35L/c_{s0}$, we obtain
${\bm\omega}_{\rm min} = (2.61 - 0.348i, 3.54 - 0.364i, 4.13 - 0.449i)c_{s0}/L$.
In the following, we refer to $u({\bm\omega}_{\rm min},{\bm C}_{\rm min},v_{\infty,{\rm min}};t)$ as the ``best fitting'' function.
In fact, the waveform is well approximated by the ``best fitting'' function
for $t \agt 18L/c_{s0}$, as shown in the lower panel of Fig.~\ref{fig:fittingCS1}.
In addition, we find that ${\bm\omega}_{\rm min}$ agree with three of the boxed
quasinormal frequencies calculated from Eq.\eqref{eq:BQNM_PT}, $(\omega_{\rmb,3},\omega_{\rmb,4},\omega_{\rmb,5})=(2.47 - 0.356i, 3.34 - 0.353i, 4.15 - 0.430i)c_{s0}/L$, within 6\% accuracy.
For comparison, a superposition of the three boxed quasinormal modes is also plotted in Fig.~\ref{fig:fittingCS1}.
It is noted that the quality factor of ${\bm\omega_{\rm min}}$ is $Q = 4.44$ on average,
while that of the least-damped ordinary quasinormal mode is $Q_0 = 1.41$ 
(in the P\"{o}schl-Teller approximation).
It is quite surprising that a weakly-reflecting wall with the reflectance $|{\cal R}|^2 \sim 1 \%$ enhances the quality
factor of the ringing by $\sim 300\%$.

We carry out type I' simulations for different $x_c$ and obtain ${\bm\omega}_{\rm min}$ with the same
analysis as above, whose relults are plotted in Fig.~\ref{fig:CSdistanceReIm}. 
In the analysis, we need to increase the number $j_{\rm max}$ of the damped sinusoids with increasing $x_c$,
since the number of the boxed quasinormal modes with small $|\Im(\omega_\rmb)|$ also increases. 
It is found that the observed frequencies corresponds to those
of the boxed quasinormal modes with $\Re(\omega_{\rmb,m})$ close to $\Re(\omega_0)$, implying that the
boxed quasinormal modes which are ``close'' (in the frequency space) to the least-damped ordinary quasinormal mode are strongly excited. This is narural since the boxed quasinormal modes
are excited by the quasinormal ringing at the early time. Roughly speaking, the late-time ringing inherits the central frequency from the quasinormal ringing at the early time, 
while its damping rate is greatly supressed due to the reflection at the boundary.
This is the main reason why the quality factor ($\propto$ the central frequency divided by the damping rate) is greatly enhanced by the reflective boundary.
 
\begin{figure}
\includegraphics[width=8.0cm]{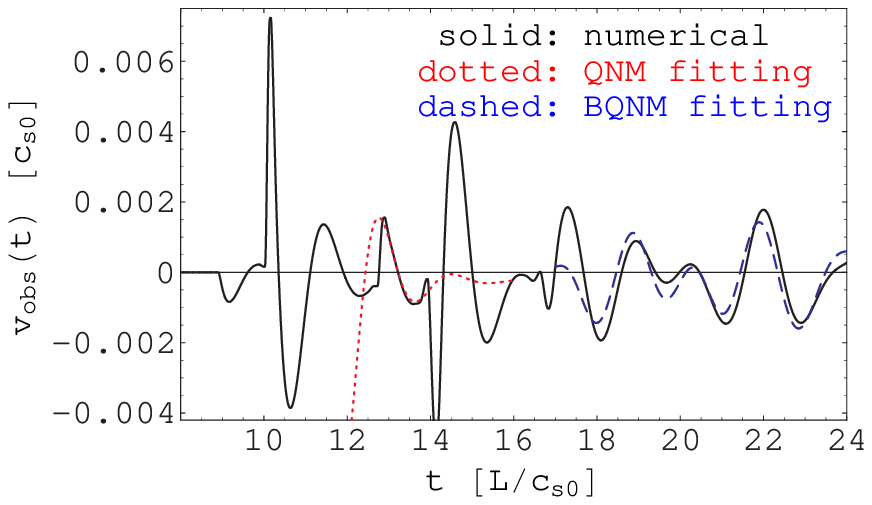}
\includegraphics[width=8.0cm]{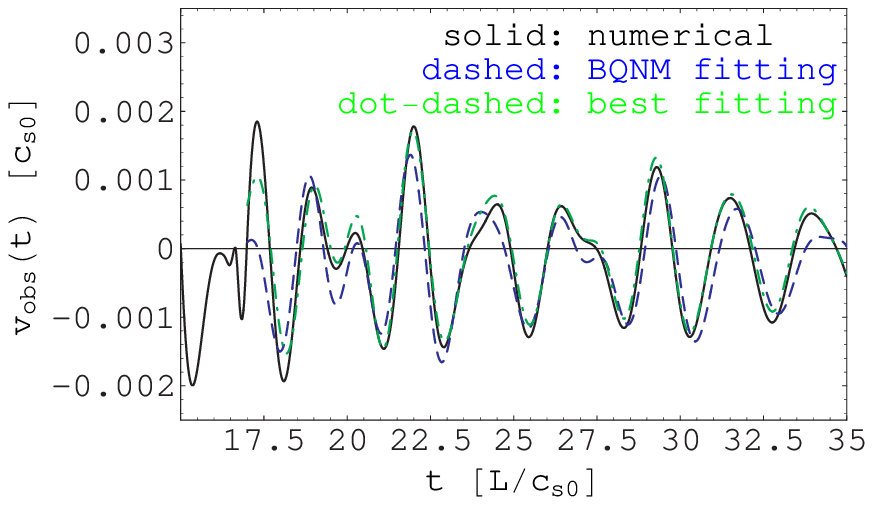}
\vspace{-3mm}
\caption{
Numerical waveform $v_{\rm obs}(t) = v(t,2.0L)$ (solid) for an early time (upper panel) and a late time (lower panel), obtained from the type I' simulation with $(\alpha,\beta,{\cal R},x_c)=(2,0.01,1.0,4.0L)$.
}
\label{fig:fittingCS2}
\end{figure}
Up to here, we have dealt with the effect of a weakly-reflecting boundary (${\cal R} \ll 1$)
and shown that the our formula \eqref{eq:BQNM_PT} gives the boxed quasinormal frequencies 
in a good accuracy for that case.
Now we show that the formula \eqref{eq:BQNM_PT} is also valid for a perfectly-reflecting boundary
(${\cal R}=1$).
Fig.~\ref{fig:fittingCS1} shows the numerical waveform of $v$ at $x_{\rm obs}=2.0L$ obtained from a type I' simulation with $(\alpha,\beta,{\cal R},x_c)=(2,0.01,1.0,4.0L)$.
We carry out for this waveform the same analysis as above, and obtain ${\bm\omega}_{\rm min} = (2.59 - 0.00546i,3.45-0.0430i,4.28-0.102i)c_{s0}/L$.
These values agree with three of the boxed quasinormal frequencies calculated from Eq.~\eqref{eq:BQNM_PT}, $(\omega_{\rmb,3},\omega_{\rmb,4},\omega_{\rmb,5})=(2.48-0.00457i,3.36-0.0302i,4.20-0.114i)c_{s0}/L$.
The averaged quality factor of ${\bm\omega_{\rm min}}$ is $Q = 99.4$, which is far greater than $Q_0$.
\section{Conclusion}
We have studied the quasinormal modes of acoustic black holes in Laval nozzles 
both semianalytically and numerically.
We have found that the quasinormal modes are actually excited in response to perturbations to
the transonic flow, but also found that the quality factor is much smaller than 10, i.e., the quasinormal ringing damps within only a few oscillation cycles.
In an actual experiment, however, the purely-outgoing boundary condition will not be satisfied
due to the reflection at the end of the apparatus,
and the observed ringing will be expressed as a superposition of ``boxed" quasinormal modes, not of the ordinary quasinormal modes.
The quality factor of the late-time ringing can exceed that of the least-damped ordinary quasinormal mode, 
since the central frequency is basically inherited from the early-time, ordinary quasinormal ringing
while the damping rate is greatly suppressed by the reflective boundary wall.
Therefore, if the inner wall at the uppstream end of the experimental apparatus is well arranged  so that it reflects sound waves as perfectly as possible, the detection of the quasinormal ringing in a laboratory will be done quite efficiently.

In this study, we have dealt with quasi-one-dimensional transonic flows only.
More generally, it is possible to consider a flow with a nonvanishing azimuthal velocity component,
an acoustic analogue of a rotating black hole.
It is known that the quality factor of the least-damped quasinormal mode is larger
for rotating black holes than for non-rotating.
Hence, the quality factor of the quasinormal modes for a ``rotating" acoustic black hole
(transonic flow with azimuthal rotation) may be large than that for ``non-rotating'' ones.
Also, it is considered that a reflective shell that encloses a rotating black hole can explode due to 
the super-radiant scattering by the black hole; such a system is well-known as a ``black hole bomb''~\cite{PressTeukolski}.
Similarly, a rotating acoustic black hole with a reflective boundary wall is expected to
behave as an ``acoustic black hole bomb"~\cite{Berti04}.
We are planning to perform numerical simulations for rotating acoustic black holes in Laval nozzles and investigate these issues in the future works.

\acknowledgments
We are grateful to S. Inutsuka for providing us his Riemann solver code and introducing us to computational fluid dynamics. Some of the numerical simulation were carried out on SX8 at YITP in Kyoto University. This work was supported by JSPS Grant-in-Aid for Exploratory Reserach No.~19656056.

\end{document}